\documentclass[aps,prd]{revtex4}

\def\beq{\begin{equation}}
\def\eeq{\end{equation}}
\def\beqa{\begin{eqnarray}}
\def\eeqa{\end{eqnarray}}

\def\zb{\beta}
\def\lsim{\mathrel{\raise.3ex\hbox{$<$\kern-.75em\lower1ex\hbox{$\sim$}}} }
\def\gsim{\mathrel{\raise.3ex\hbox{$>$\kern-.75em\lower1ex\hbox{$\sim$}}} }

\begin{document}
\thispagestyle{empty}

\begin{flushright}
NCU-HEP-k012  \\
Nov 2003
\end{flushright}

\vspace*{.5in}

\begin{center}
{\bf  \Large Flavor and Little Higgs $^*$}\\
\vspace*{.5in}
{\bf  Otto C.W. Kong}\\[.05in]
{\it Department of Physics, National Central University, Chung-li, TAIWAN 32054 \\
E-mail: otto@phy.ncu.edu.tw}

\vspace*{.8in}
{\bf Abstract}\\[.2in]
We discuss the proper starting point to look into flavor physics under the 
perspective of solving the hierarchy problem with the little Higgs mechanism 
--- the construction of anomaly free fermionic spectra of the effective theory of 
extended electroweak gauge symmetry around the TeV scale. Anomaly
cancellations among the three SM families, plus extra fermions,  are typically 
needed, giving family non-universal flavor structures. 
Such a completed and consistent fermionic spectrum supplements
a scalar sector little Higgs model, turning the latter into a
realistic model the phenomenological viability of which can 
then be investigated. The implications for flavor physics are
specially interesting. We give more elaborated discussion here on our
basic perspective, trying to clarify on issues that are in our opinion
under-appreciated. We also add a brief comment of little Higgs versus
supersymmetry.
\end{center}

\vfill
\noindent --------------- \\
$^\star$ Invited Talk presented  at ICFP 2003 (Oct 6--11), KIAS, Korea\\
 --- submission for the proceedings. 
 
\clearpage
\addtocounter{page}{-1}

\title{Flavor and Little Higgs\\[.2in]
 }
\author{ Otto C. W. Kong}
\affiliation
{
Department of Physics, National Central University, Chung-li, Taiwan 32054\\
}

\begin{abstract}
We discuss the proper starting point to look into flavor physics under the 
perspective of solving the hierarchy problem with the little Higgs mechanism 
--- the construction of anomaly free fermionic spectra of the effective theory of 
extended electroweak gauge symmetry around the TeV scale. Anomaly
cancellations among the three SM families, plus extra fermions,  are typically 
needed, giving family non-universal flavor structures. 
Such a completed and consistent fermionic spectrum supplements
a scalar sector little Higgs model, turning the latter into a
realistic model the phenomenological viability of which can 
then be investigated. The implications for flavor physics are
specially interesting. We give more elaborated discussion here on our
basic perspective, trying to clarify on issues that are in our opinion
under-appreciated. We also add a brief comment of little Higgs versus
supersymmetry.
\end{abstract}
\maketitle

\section{Introduction}
This is second talk on little Higgs in the conference program. We will take
advantage of the first talk from T.Han\cite{han}, in which the little Higgs idea
has been reviewed. Instead of repeating what has been said there, we would
rather focus on our main concern --- the construction of an anomaly free
fermionic spectrum as a necessary completion of a little Higgs model and
the proper starting point to think about flavor physics under the model. To set
a clear perspective, however, we supplement a brief comment of the apparent
difference between the main physics of the first talk and that to be addressed
here.

The simplest way to look at the question is to say that we are talking about
different (kinds of) little Higgs models, namely, we focus on models from
Kaplan and Schmaltz\cite{KS} with $SU(N)_L\times U(1)_X$ ($N=3$
and 4) extended electroweak (gauge) symmetries. For instance, for the $N=3$ case,
we have actually a scalar (Higgs) sector of a $[SU(3)]^2/[SU(2)]^2$ nonlinear 
sigma model, as versus that of the $SU(5)/SO(5)$, model\cite{ltH} on which the first
talk is mainly based. In our opinion, the models discussed here actually have
structures that are more carefully worked out, especially after our effort on
which the current talk is mainly based\cite{009,010}. However, we certainly
believe that the issues addressed here have relevance for generic little Higgs
model building as well as little Higgs flavor physics. Such issues  have not
been well addressed by earlier authors on the subject, who might be focusing
more on establishing and illustrating the little Higgs mechanism rather 
than furnishing complete particle physics models. For example, the full fermionic
content of all the multiplets under the global as well as extended gauge
symmetries of the $SU(5)/SO(5)$ model has not been presented. To elaborate more, 
only one extra fermion, the heavy top quark $T$ needed to cancel the quadratic 
divergence (of the contribution to Higgs mass) from the SM $t$ quark at 1-loop 
is discussed. It is taken to be vectorlike under the SM gauge symmetry. But 
there should be a lot more to the story. The $T$ quark, or rather its chiral
components, has to join the SM chiral components to form multiplet(s) under
the extended symmetries. There would then be other extra fermions from the
full multiplet(s). It is also very unlikely there one can arrange the little Higgs
mechanism with the $T$ quark, as well as all the other extra fermions, vectorlike 
under the  full gauge symmetry of the model. Neither is that  exactly a
desirable feature. If the extra states are fully vectorlike, instead of forming
vectorlike pairs only after the gauge symmetry is broken to that of the SM, there
is no reason for such states, including the $T$, to have masses below the
10-50 TeV cutoff scale. For the kind of models we discussed here below, we will
spell out the complete multiplets necessary to house all the fermions --- SM ones 
plus $T$ as a minimal list. For each particular model, the list is chiral to begin with 
while yielding the SM chiral list plus vectorlike states including the $T$ as SM gauge 
symmetry becomes the only surviving gauge symmetry. The latter is exactly what is
required to be a consistent extended symmetry model embedding the SM. Furthermore, 
we will see that such a list is essentially dictated by the particular realization
of the little Higgs mechanism and the condition that chiral fermionic spectrum be 
gauge anomaly free. The gauge anomaly cancellation conditions constitute what 
we believe to be the best understanding we have on the question of {\it why there is
what there is}. And, in our opinion, it is apparently much under-appreciated by
many physicists.

Han and collaborators has performed some interesting phenomenological studies of 
the $SU(5)/SO(5)$ little Higgs model\cite{han}. Most parts of those results are
valid, as they are independent of the complete fermionic spectrum, {\it i.e.} when
one assumes that such a consistent spectrum can be obtained, and any extra fermion
to be included has a mass very close to the cutoff and/or has no significant coupling
to the SM spectrum. However, looking at the little Higgs models from a model-building
point of view, one should wonder if we can do a more complete job. On the other hand,
if our concern is flavor physics, the detailed representation assignment of the SM fermions
themselves under the extended symmetries would have strong implications on the
flavor structure of a model. Hence, it becomes an unavoidable subject. Not to say that 
the extra fermions may also play some role. What we are doing here is only a small 
starting step in the direction --- a direction that certainly deserves more attention if
we want to consider any little Higgs model as a realistic particle physics model though.

\section{The \boldmath $SU(3)_L\times U(1)_X$ little Higgs model}
Let us first give a summary of the particular realization of little Higgs mechanism our
discussion explicitly based on, the model(s) from Kaplan and Schmaltz\cite{KS}. The
latter paper, as well as Ref.\cite{S}, presents a simple group theoretical approach
to little Higgs model building, which might look more transparent and easier to
follow for beginners. The construction focuses only on the TeV scale effective field
theory, independent of its strong couplings parent theory or so-called UV-completion. It 
is the same bottom-up model building perspective that we adopted here, only pushing
it all the way to the end. We are looking into the structure of the little Higgs model
as a {\it complete} particle physics model --- effective field theory, living between
the electroweak scale and the scale of the yet unknown UV-completion. We ask
for a consistent and comprehensive description of its physics content. The 
phenomenological viability of the model may then be carefully studied and
tested experimentally.

Kaplan and Schmaltz considered a scalar sector with two (anti)triplets 
$\Phi_{\!\scriptscriptstyle 1}$ and $\Phi_{\!\scriptscriptstyle 2}$
each taken as carrying its own  $SU(3)$ symmetry. The two (anti)triplets are assigned
VEVs aligned within a combination of the two  $SU(3)$'s that is gauged together 
with an extra $U(1)_X$. The two sets of five Goldstone states from the two 
$SU(3) \to SU(2)$ symmetry-breakings then consists of an unphysical set to be 
eaten by the extra ``electroweak" gauge bosons in the $SU(3)_L\times U(1)_X$
beyond their SM electroweak cousins. The orthogonal set contains 
pseudo-Nambu-Goldstone bosons (PNGBs) under the remaining part of the symmetries 
which are global and really (can be) only  approximate in the full Lagrangian.
The PNGBs include the SM Higgs doublet and an extra singlet scalar. 
Explicitly, there is the following nonlinear sigma model description :
\beq
\Phi_1=e^{i \theta/f}
\left( \begin{array}{ccc}
0  & 0 & f \end{array} \right)\! \qquad\qquad
\Phi_2=e^{-i \theta/ f}
\left( \begin{array}{ccc}
0  & 0 & f \end{array} \right)\! \;,
\eeq
\beq
\theta={1 \over \sqrt{2}}
\left( \begin{array}{cc} 
\!\!\begin{array}{ll} 0 & 0 \\ 0 & 0 \end{array} 
& \!\!h^\dagger \\ h & \!\!0 \end{array} \right)
+{\eta \over 4} 
\left( \begin{array}{rrr} 1&0 &0 \\0 &1&0\\ 0& 0&\!\!\!-2 \end{array}
\right)  \;.
\eeq

To take care of the top sector contribution to Higgs mass correction at 1-loop,
the extra $T$ quark is introduced with a chiral component living inside a
triplet of the extended electroweak gauge symmetry. This is exactly the
triplet that contains the $(t,b)$ SM doublet, hence linking $T$ to $t$. We
have Yukawa terms involving the two top-like quarks given as
\beqa
{\mathcal L}_{top} &=& y_{\!\scriptscriptstyle 1}\,\bar{t}'_a\,
{\Phi_{\!\scriptscriptstyle 1} \, Q^a}   +    y_{\!\scriptscriptstyle 2}\,\bar{T}'_a\,
{\Phi_{\!\scriptscriptstyle 2} \, Q^a}
\nonumber \\  &=& \label{tY}
f\,(y_{\!\scriptscriptstyle 1}\,\bar{t}'+  y_{\!\scriptscriptstyle 2}\,\bar{T}')\, T
+ \frac{i}{\sqrt{2}} \, (y_{\!\scriptscriptstyle 1}\,\bar{t}'-  y_{\!\scriptscriptstyle 2}\,\bar{T}')\,
{h\left( \begin{array}{c} t \\ b \end{array} \right)} + \cdots \;,
\eeqa
where $Q^a=(T^a, t^a, b^a)^{\!\scriptscriptstyle T}$,  and
$\bar{t}'_a$, and  $\bar{T}'_a$ denote colored singlet states. For details on the
functioning of the little Higgs mechanism, we refer readers to the original paper\cite{KS}.
From our discussion at the beginning, we are led to the question of how to complete
the group theoretical description of the full fermion spectrum. All SM doublets would
have to come from nontrivial representations of the $SU(3)_L$. The first point to note
here is that simply repeating the structure here with quarks for the lighter two
families does not work. The simple embedding produces $SU(3)_L$ gauge
anomaly rendering the model inconsistent. Nor is such an embedding desirable,
as we will show below.

\section{Gauge anomaly Vs Flavor structure --- a detour} 
To set a good reference background for the problem at hand, we take a detour here
to re-examine the SM spectrum and the questions of flavor structure. The spectrum
of SM fermion in one family is like perfection, essentially dictated by gauge
anomaly cancellation conditions.  And there is the million-dollar question, or 
fundamental question of flavor physics --- Why three families ? To illustrate the point
of view, we recall our earlier argument\cite{unc67}. Assuming that there
exist a minimal multiplet carrying nontrivial quantum numbers of each of the
component gauge groups, one can obtain the one-family SM spectrum as the
unique solution by asking for the minimal consistent set of chiral states. A vectorlike
set is trivial but not as interesting. Only chiral states are protected from heavy
gauge invariant masses. 

The above suggested derivation of the one-family SM spectrum goes as follow.
We are essentially starting with a quark doublet,  with arbitrary hypercharge
normalization. The two $SU(3)_C$ triplets require two antitriplets to cancel
the anomaly. Insisting on the chiral spectrum means taking two quark singlets
here, with hypercharges still to be specified. Now, $SU(2)_L$ is real, but has
a global anomaly. Cancellation requires an even number of doublet, so at least
one more beyond the three colored components in the quark doublet. There
are still four anomaly cancellation conditions to take care of. They are the
$[SU(3)_C]^2 U(1)_Y$, $[SU(2)_L]^2 U(1)_Y$, $[grav]^2 U(1)_Y$, and
$[U(1)_Y]^3$. We are however left with three relative hypercharges to fit the
four equations actually without a possible solution. A rescue comes from simply
adding a $U(1)_Y$-charged singlet. But the four equation for four unknown
setting is misleading. The $[U(1)_Y]^3$ anomaly cancellation equation is cubic
in all the charges, with no rational solution guaranteed. The SM solution may
actually be considered a beautiful surprise.

We would also like to take the opportunity here to briefly sketch the next step taken
in Ref.\cite{unc67}, to further illustrate our perspective. The results there also may
be considered a worthy comparison with our little Higgs motivated flavor/family
spectrum presented below, from the point of view of the origin of the three families.
The major goal of Ref.\cite{unc67} is to use a similar structure with an extended
symmetry to obtain the three families. For example, one can start with some
$SU(4)\times SU(3)\times SU(2)\times U(1)$ gauge symmetry and try to obtain
the minimal chiral spectrum contain a $(4,3,2)$ multiplet --- the simplest one
with nontrivial quantum number under all component groups. Having a consistent
solution is not enough though. In order for the spectrum be of interest, we ask the 
spectrum to yield  the chiral spectrum of three SM families plus a set of vectorlike
states under a feasible spontaneous symmetry breaking scenario, {\it i.e.} when
the gauge symmetry is broken to that of the SM. Ref.\cite{unc67} has only
partial success. A consistent group theoretical SM embedding could be obtained
with a slight addition to the minimal chiral spectrum obtained from anomaly
cancellation considerations alone. We give an example in Table I.

To conclude the section, we emphasize that anomaly cancellation conditions
play a role of paramount importance in constraining the fermionic spectrum.
Having a complete model with an extended gauge symmetry above the electroweak 
scale successfully, even though only group theoretically speaking, embedding
the SM states is not at all trivial. Types of consistent models of the kind in the
literature are quite limited. However, such exercises might, and have been,
used as a way to probe the most fundamental question of flavor physics --- why 
3 families? We return to the little Higgs model(s) below, to get
exactly such a job finished. We will then see how the requirement for an
anomaly free chiral fermion spectrum dictates a specific flavor structure to
a particular model. Readers are invited to consider if the model, or its
fermion spectrum, obtained is as esthetically appealing as the (one-family) SM 
or the models obtained in Ref.\cite{unc67}. We do believe that such
esthetic concerns may have some relevance in fundamental physics. After all,
it cannot be emphasized enough that the anomaly cancellation requirement is
the best, and might be only, theoretical tool we have up to the moment to address
the question of why there is what there is in terms of the fermionic sector. 


\begin{center}
\small\noindent
Table I. A $SU(4)_A\otimes SU(3)_C\otimes SU(2)_L\otimes U(1)_X$
anomaly free chiral fermion spectrum with three SM families.\\[.05in]
\begin{tabular}{|c|c|r|r|r|r|r|cc|}
\hline\hline
 multiplets & $X$ &
\multicolumn{5}{|c|}{Gauge anomalies} &  \multicolumn{2}{|c|}{$U(1)_Y$ states } \\ 
\hline 
&  &	$U(1)$-grav. & $[SU(4)]^2U(1)$	& $[SU(3)]^2U(1)$ &  $[SU(2)]^2U(1)$ &	$[U(1)]^3$ &				& \\
\hline								  		
${\bf (4,3,2)}$		&	{\bf 1}  &	24  &	6  &	8 &	12 &	24	& 	3\ {\bf 1}($Q$)	   	& {\bf -5}($Q^{'}$) \\
${\bf (\bar{4},\bar{3},1)}$ &	{\bf 5}  &	60  &	15 & 	20 &	 &	1500	& 	3\ {\bf -4}($\bar{u}$)	& {\bf 2}($\bar{d}$) \\
${\bf (\bar{4},1,2)}$	&	{\bf 3}  &	24  &	6  &	&	12 &	216	& 	3\ {\bf -3}($L$)	& {\bf 3}($\bar{L}$) \\
${\bf (\bar{4},1,1)}$	&	{\bf 9}  &	36  &	9  &	&	 &	2916	& 	3\ {\bf -6}($\bar{E}$)	& {\bf 0}($N$) \\
${\bf (6,1,1)}$		&	{\bf -18} &	-108 &	-36 &	&	 &	-34992	& 	3\ {\bf 6}($E$)	& 3\ {\bf 12}($S$) \\ \hline
${\bf (1,\bar{3},2)}$	&	{\bf -10} &	-60 &	&	-20 &	-30 &	-6000	& 	\multicolumn{2}{|c|}{{\bf 5}($\bar{Q}^{'}$)} \\
${\bf (1,\bar{3},1)}$	&	{\bf -4} & 	-12 &	&	-4  &	 &	-192	& 	\multicolumn{2}{|c|}{{\bf 2}($\bar{d}$)} \\
${\bf (1,\bar{3},1)}$	&	{\bf -4} &	-12 &	&	-4  &	 &	-192	& 	\multicolumn{2}{|c|}{{\bf 2}($\bar{d}$)} \\ \hline
${\bf (1,1,2)}$		&	{\bf 6}  &	12  &	&	&	6  &	432	&	\multicolumn{2}{|c|}{{\bf -3}($L$)} \\
$3\ {\bf (1,1,1)}$ 	&	{\bf 24} &	72  &	&	&	&	41472	&	\multicolumn{2}{|c|}{3\ {\bf -12}($\bar{S}$)} \\ 
$3\ {\bf (1,1,1)}$ 	&	{\bf -12} &	-36 &	&	&	&	-5184	&	\multicolumn{2}{|c|}{3\ {\bf 6}($E$)} \\	\hline
\multicolumn{2}{|r|}{\it Total}    	&	0   &	0  &	0  &	0  & 	0	& 				& \\	
\hline\hline
\end{tabular}
\end{center}
\normalsize

\section{Anomaly free \boldmath $SU(N)_L\times U(1)_X$ spectra}
In our recent papers\cite{009,010}, we have illustrated that an anomaly free fermionic spectrum
with consistent embedding of the three family SM fermions can be obtained. In fact, we have given 
a recipe for an infinite number of similar fermionic spectra of with an  $SU(N)_L\times U(1)_X$ 
extended electroweak symmetry for any $N>2$. The basic feature of such a spectrum is to have
$t$-$b$ quark SM doublet embedded into a fundamental representation of $SU(N)_L$ while
the lighter two quark doublets embedded into anti-fundamental, as inspired by Ref.\cite{PF}.
It exploits the fact that
\[
N_c = N_f \;;
\]
namely, the number of SM families (of fermions) $N_f$ happens to coincide with the number
of colors. Indeed such a spectrum  may be considered rather as the reason behind the number 
of SM family. We emphasize, again, that the need for gauge anomaly cancellation, for the SM 
or otherwise, is the best we have in addressing the question if why there is what there is --- 
a perspective we try to illustrate in the last section. The $1+2$ quark family embedding has 
net $SU(N)_L$ anomaly to be canceled exactly by embedding the three leptonic doublets into
fundamentals. The simplest way to complete the spectrum with consistent SM embedding
is to put anything else in singlets, with just enough of them to produce vector-like pairs for
all single fermion states at the QED and QCD level. Neutrinos may be sort of exceptions to the
last statement. Carrying no conserved gauge quantum number, they can be considered
self-conjugate. When checking the other gauge anomalies of such a spectrum, it looks like a 
miracle upon a first glance. There are prefect cancellations for all, as illustrated by the 
$N=3$ example given in Table II.  Two other example spectra given in Table III below, 
in which we skip the explicit presentation of the gauge anomalies. Readers can easily 
checked that the cancellations do work. In fact, a careful checking of the algebra\cite{010}
would show that the recipe as outlined here always works for any $N$. Looking at the 
possible ways of obtaining the hypercharge $U(1)$ out of the $N-1$ $U(1)$s inside the
$SU(N)_L$ and the still extra $U(1)$ of $U(1)_X$, one will see that there is exactly a $N-2$
parameter degree of freedom and choosing such a model spectrum. That can be taken as
the freedom for choosing the electric charges of the $N-2$ extra quark states within the
same $SU(N)_L$ multiplet as the $t$-$b$ doublet. 

\noindent\begin{center}
\small
Table II : The $SU(3)_C\times SU(3)_L\times U(1)_X$ spectrum with 
little Higgs. Electroweak doublets are put in [.]'s. \\[.05in]
\begin{tabular}{|c|r|r|r|r|r|cc|}
\hline\hline
{fermion multiplets}  & \multicolumn{5}{|c|}{Gauge anomalies} &  \multicolumn{2}{|c|}{$U(1)_Y$ states } \\ 
\hline 
    &  	 $U(1)$-grav. & $[SU(3)_L]^3$	& $[SU(3)_L]^2U(1)$ &  $[SU(3)_C]^2U(1)$ &	$[U(1)]^3$ &				& \\
\hline								  		
${\bf (3_{\scriptscriptstyle C},3_{\scriptscriptstyle L} ,\frac{1}{3})}$
      &    $\frac{1}{3}\times 9\times 1$ &	$3\times 1$  &	$\frac{1}{3}\times 3\times 1$ &	$\frac{1}{3}\times 3\times 1$ &  $\frac{1}{27}\times 9\times 1$	 &  ${\bf \frac{1}{6}}$[$Q$]               & ${\bf \frac{2}{3}}$($T$)    \\
2\ ${\bf ({3}_{\scriptscriptstyle C},\bar{3}_{\scriptscriptstyle L} ,0)}$
     &       0 &	$-3\times 2$ & 	$0$	 & 0 &	0     &  2\ ${\bf \frac{1}{6}}$[$Q$]               & 2\ ${\bf \frac{-1}{3}}$($D,S$)     \\
$3\ {\bf (l_{\scriptscriptstyle C} ,3_{\scriptscriptstyle L} ,\frac{-1}{3})}$
    &	 $\frac{-1}{3}\times 3\times 3$&	$1\times 3$  &	   $\frac{-1}{3}\times 3$ &	 & $\frac{-1}{27}\times 3\times 3$	           & 3\  ${\bf \frac{-1}{2}}$[$L$]          	& 3\ {\bf 0}($N$)   \\
$4\ {\bf (\bar{3}_{\scriptscriptstyle C},1_{\scriptscriptstyle L} ,\frac{-2}{3})}$
   &	$\frac{-2}{3}\times 3\times 4$&	 &	&	 $\frac{-2}{3}\times 4$ & $\frac{-8}{27}\times 3\times 4$        & \multicolumn{2}{|c|}{4\ ${\bf \frac{-2}{3}}$ ($\bar{u}, \bar{c}, \bar{t}, \bar{T}$)} 	\\
$5\ {\bf (\bar{3}_{\scriptscriptstyle C},1_{\scriptscriptstyle L} ,\frac{1}{3})}$  
   &	$\frac{1}{3}\times 3\times 5$&	&	&	 $\frac{1}{3}\times 5$ & $\frac{1}{27}\times 3\times 5$        & \multicolumn{2}{|c|}{5\ ${\bf \frac{1}{3}}$ ($\bar{d}, \bar{s}, \bar{b}, \bar{D}, \bar{S}$)}\\
$3\ {\bf (1_{\scriptscriptstyle C},1_{\scriptscriptstyle L} ,1)}$   	
    &     $1\times 3$  &	 &	&	 &	$1\times 3$           &    \multicolumn{2}{|c|}{3\ ${\bf 1}$  ($e^+, \mu^+, \tau^+$)   }       \\ 
 \hline
\multicolumn{1}{|r|}{ Total}    	&	0  &   0  &	0  &	0  & 	0	& 				& \\	
\hline\hline
\end{tabular}
\end{center}
\normalsize

We have summarized in the last paragraph the possible anomaly free fermionic spectra with
$SU(N)_L\times U(1)_X$ extended electroweak gauge symmetries. We would
certainly love to arrive at a conclusion of only one or two of such spectra are possible, instead
of having infinitely many candidates at hand. However, not all of such spectra
would be relevant to little Higgs, which is the major motivation of our study here. 
The first obvious criterion for such a fermionic spectrum to fit into a little Higgs model is
the existence of the $T$ quark --- a heavy top quark that couples to the SM Higgs in such a
way as to allow the cancellation of quadratic divergence. Following what is illustrated in
Ref.\cite{KS} [{\it cf.} Eq.(\ref{tY})], we simply ask for a $T$ quark within the $Q^a$
multiplet --- the $SU(N)_L$ multiplet containing the $t$-$b$ doublet. (Note also that in the
Tables, $Q$ stands for a SM quark doublet instead.) For the case of $N=3$, that fix the
spectrum as given in Table II as the only option. 

Naively, one would prefer to stay with a smaller $N$. The unique $N=3$ case fitted with
the Higgs sector structure introduced in Ref.\cite{KS}, however, has some problem with getting
the Higgs quartic coupling, as noted by the authors. Interesting enough, another group
published a different little Higgs model with again the $SU(3)_L\times U(1)_X$ extended
electroweak symmetry\cite{ST}. In the latter case, the gauge symmetry is embedded into a 
global $SU(9)$. It certainly looks like the fermionic spectrum could play a role in the 
$SU(9)$ model too, though we are not really ready to comment of that here.

\section{ \boldmath $SU(3)_C\times SU(4)_L\times U(1)_X$ models}
To surmount the quartic coupling barrier of the  $N=3$ case, Ref.\cite{KS}
turns to the construction of a $N=4$, with simply two $T$ quark instead of one. Our
analysis of the fermionic sector suggests the first spectrum given in Table III works as
a completion of the model. Another interesting question that arises is on the exact
identity of the fourth quark in the multiplet of $t$-$b$ and the single $T$ always required.
While we have not investigated all options in any detail, we believe the second spectrum
given in Table III would be a feasible and interesting alternative, with a heavy bottom $B$
quark instead of a second $T$. At the electroweak level, the type of $N=4$ little Higgs
model has two SM Higgs doublet. The modified Higgs sector we introduced in Ref.\cite{010}
to accompany the second  $N=4$ fermionic spectrum looks more like a two Higgs doublet
model with natural flavor conservation and a naturally large $\tan\!\zb$. The latter
are phenomenologically desirable\cite{2hdm}.

\begin{center}
\small
Table III : Fermion spectra for two  $SU(3)_C\times SU(4)_L\times U(1)_X$  little Higgs models.
\\
\begin{tabular}{|c|ccc||c|ccc|}
\hline\hline
 \multicolumn{4}{|c||}{First model spectrum}	 &  \multicolumn{4}{|c|}{Second model spectrum}			\\  \hline

				& \multicolumn{3}{|c||}{$U(1)_Y$-states}	& & \multicolumn{3}{|c|}{$U(1)_Y$-states}			\\  \hline
${\bf (3_{\scriptscriptstyle C},4_{\scriptscriptstyle L} ,\frac{5}{12})}$                 &  ${\bf \frac{1}{6}}$[$Q$]               &2\ ${\bf \frac{2}{3}}$($T,T'$)  &      				&
${\bf (3_{\scriptscriptstyle C},4_{\scriptscriptstyle L} ,\frac{1}{6})}$                 &  ${\bf \frac{1}{6}}$[$Q$]               & ${\bf \frac{2}{3}}$($T$)  &      ${\bf \frac{-1}{3}}$($B$) 	\\
2\ ${\bf ({3}_{\scriptscriptstyle C},\bar{4}_{\scriptscriptstyle L} ,\frac{-1}{12})}$     &  2\ ${\bf \frac{1}{6}}$[2\ $Q$]               & 4\ ${\bf \frac{-1}{3}}$($D,D',S,S'$)     &  &
2\ ${\bf ({3}_{\scriptscriptstyle C},\bar{4}_{\scriptscriptstyle L} ,\frac{1}{6})}$     &  2\ ${\bf \frac{1}{6}}$[2\ $Q$]               & 2\ ${\bf \frac{-1}{3}}$($D,S$)     & 2\ ${\bf \frac{2}{3}}$($U,C$) \\
$3\ {\bf (l_{\scriptscriptstyle C} ,4_{\scriptscriptstyle L} ,\frac{-1}{4})}$           & 3\  ${\bf \frac{-1}{2}}$[3\ $L$]          	& 6\ {\bf 0}(6\ $N$) 		 &	&
$3\ {\bf (l_{\scriptscriptstyle C} ,4_{\scriptscriptstyle L} ,\frac{-1}{2})}$           & 3\  ${\bf \frac{-1}{2}}$[3\ $L$]          	& 3\ {\bf 0}(3\ $N$) 		 & 3\ {\bf -1}(3\ $E^-$)  \\
$5\ {\bf (\bar{3}_{\scriptscriptstyle C},1_{\scriptscriptstyle L} ,\frac{-2}{3})}$        & \multicolumn{2}{|c}{5\ ${\bf \frac{-2}{3}}$ ($\bar{u}, \bar{c}, \bar{t}, \bar{T}, \bar{T'}$)}     & 	  			&
$6\ {\bf (\bar{3}_{\scriptscriptstyle C},1_{\scriptscriptstyle L} ,\frac{-2}{3})}$        & \multicolumn{3}{|c|}{6\ ${\bf \frac{-2}{3}}$ ($\bar{u}, \bar{c}, \bar{t}, \bar{U}, \bar{C}, \bar{T}$)}     	           	\\
$7\ {\bf (\bar{3}_{\scriptscriptstyle C},1_{\scriptscriptstyle L} ,\frac{1}{3})}$        & \multicolumn{2}{|c}{7\ ${\bf \frac{1}{3}}$ ($\bar{d}, \bar{s}, \bar{b}, \bar{D}, \bar{D'}, \bar{S}, \bar{S'}$)} 	&               			&
$6\ {\bf (\bar{3}_{\scriptscriptstyle C},1_{\scriptscriptstyle L} ,\frac{1}{3})}$        & \multicolumn{3}{|c|}{6\ ${\bf \frac{1}{3}}$ ($\bar{d}, \bar{s}, \bar{b}, \bar{D}, \bar{S},\bar{B}$)} 	               	\\
$3\ {\bf (1_{\scriptscriptstyle C},1_{\scriptscriptstyle L} ,1)}$              &    \multicolumn{2}{|c}{3\ ${\bf 1}$  ($e^+, \mu^+, \tau^+$)   }      	 				&						&
$3\ {\bf (1_{\scriptscriptstyle C},1_{\scriptscriptstyle L} ,1)}$              &    \multicolumn{2}{|c}{3\ ${\bf 1}$  ($e^+, \mu^+, \tau^+$)   }      	 & 3\ {\bf 1}(3\ $E^+$)   	\\
\hline\hline
\end{tabular}
\end{center}
\normalsize

\section{Little Higgs Vs Supersymmetry}
From the pure theoretical side, such models seem to hold great promise.
Like supersymmetry (SUSY), we have consistent complete model(s) of TeV scale physics 
successful handling the notorious hierarchy problem. While the question of family
triplication of SM fermions is not at all touch by supersymmetrizing the SM, the bosonic
symmetry scheme of little Higgs adds fermions instead of sfermions, complicating
the issues of gauge anomaly cancellation. We see here that solving the latter problem
sheds some light on the number of family problem, at least for the class of little Higgs
models we focus on here. A similar construction of anomaly
free spectra does not work for the number of family being not three (as the number of 
color). While the supersymmetric SM is quite unique, there are already many little Higgs 
models on the market. Some models may not share much common structure with
the ones discussed here, though the basic anomaly cancellation concern is always
relevant. SUSY adds to the phenomenological flavor problems. Flavor changing
neutral current (FCNC) constraints impose stringent conditions on the soft SUSY
breaking parameters. FCNC constraints are certainly no less demanding on little Higgs
models. Precision electroweak constraints are generally more difficult for the latter\cite{han}.
However, it is only with completed models as those exhibited here that one can launch
a careful study of such constraints and hence the phenomenological viability of the
models concerned. The minimal SUSY model keeps the SM neutrinos massless.
Incorporating R-parity violation is arguably natural, at the expense of introducing
many, many new couplings\cite{0045}. Our little Higgs fermionic spectra above all
contain extra (singlet) neutrino states and hence the possibility of fixing the 
experimentally required neutrino properties at the TeV scale too.

\section{Some Implications to flavor physics}
First of all, the full quantum numbers characterizing a fermion multiplet will dictate
what couplings it can have within the model. The gauge quantum numbers dictate the 
admissible gauge invariant couplings in the model Lagrangian. The latter is of course
the very starting point to look into the phenomenology of the fermionic states involved,
including those of the SM. The only other important issue here is the quantum numbers
under the global symmetries. However, as the latter are only approximate symmetries and 
have no constraints like anomaly cancellations, there are more ambiguities and rooms to
play around with. Let us illustrate some possible features with the models discussed.

Let us first restrict ourselves to the $SU(3)_L\times U(1)_X$ model originally introduced
in Ref.\cite{KS}. We have the full list of fermion gauge quantum numbers as given in Table II. 
Next, we have scalar multiplets with gauge and global quantum numbers fixed by the 
requirement of the little Higgs mechanism itself [{\it cf.} discussion on Eq.(\ref{tY})]. 
The minimal, and certainly safe, strategy is to consider no other scalar multiplets. 
Gauge quantum numbers alone say that only the following direct Yukawa couplings
are admissible :
\[
1_{\!\scriptscriptstyle L} \, \Phi_i \, {3}_{\!\scriptscriptstyle L}  
\qquad\mbox{and}\qquad
1_{\!\scriptscriptstyle L} \, \Phi_i^\dag \, \bar{3}_{\!\scriptscriptstyle L} \;.
\]
Only the $t$ and $T$ quarks come from representations of the right $X$-charges to 
have a coupling of the first from; while for the second, only the $d$ and $s$ quarks fit in.
At the next level, we have
\[
1_{\!\scriptscriptstyle L} \, \Phi_i^\dag \, \Phi_j^\dag \, {3}_{\!\scriptscriptstyle L}  
\qquad\mbox{and}\qquad
1_{\!\scriptscriptstyle L} \, \Phi_i \, \Phi_j \,\bar{3}_{\!\scriptscriptstyle L} \;.
\]
The first admits $b$ and $B$ Yukawas, as well as those for the charged leptons; the 
second admits those for $u$ and $c$. The global symmetry requirement for the `top' Yukawa
naturally admits the above `bottom' Yukawa, the higher dimensionality of which might be
a source of the smaller numerical coupling of the latter case. The 
$\bar{3}_{\!\scriptscriptstyle L}$s containing the $d$ and $s$ quark would have to bare
extra global quantum number constraints to push for high dimensional Yukawa couplings.
Such coupling suppressions have effects also on the extra $D$ and $S$ quarks, which
are likely to have strong implications to $b$ physics then.

Finally, we comment on the $SU(4)_L\times U(1)_X$ model with the second spectrum
given in Table III. The model doubles the list of SM fermions, in much the same way as SUSY
double the particle spectrum (with different spin though). The modified little Higgs part
has two pairs of  aligned-VEV $\Phi$'s with different $X$-charges to couple to the top
and bottom sector separately. The quadratic divergence cancellation then works for the 
$t$-$T$ pair as well as for the $b$-$B$ pair in similar fashion. Large bottom Yukawa is
naturally admitted. Without further global symmetry constraints, the 'top-Higgs' couples
directly to $d$- and $s$-sectors while 'bottom-Higgs' to $u$-and $c$-sectors. Flavor physics
would be very different from that of the $SU(3)_L$ model above or that of the other $SU(4)_L$
model also given in Table III. The latter is expected to be quite similar to the $SU(3)_L$
case. More detailed analysis of such models have to start with feasible complete global 
quantum number assignment to all multiplets. 

The bottom line here is that sensible discussion of flavor physics of a little Higgs model
is not possible before the full fermion spectrum is spelt out. The latter is constrained
by gauge anomaly cancellation. We exhibit some completed models here on which
detailed flavor physics still have to be studied. Building models of the kind, and studying
their phenomenology in details should be a worthy endeavor.

Our work is partially supported by the National Science Council of Taiwan, under
grant number NSC 91-2112-M-008-044.

\end{document}